\documentclass[conference]{IEEEtran}
\IEEEoverridecommandlockouts
\usepackage{cite}
\usepackage{amsmath,amssymb,amsfonts}
\usepackage{algorithmic}
\usepackage{graphicx}
\usepackage{textcomp}
\usepackage{xcolor}
\usepackage{braket}
\usepackage[FIGTOPCAP]{subfigure}
\def\BibTeX{{\rm B\kern-.05em{\sc i\kern-.025em b}\kern-.08em
    T\kern-.1667em\lower.7ex\hbox{E}\kern-.125emX}}

\newcommand{\R}{\mathbb R}
\newcommand{\Z}{\mathbb Z}

\definecolor{refblue}{RGB}{102, 102, 153}
\definecolor{stringgreen}{RGB}{80, 107, 65}
\definecolor{keywordblue}{RGB}{64, 89, 245}
\definecolor{commentbrown}{RGB}{59, 35, 0}

\usepackage{listings}
\lstdefinestyle{lststyle}{
  commentstyle=\color{commentbrown},
  keywordstyle=\color{keywordblue},
  numberstyle=\tiny\color{gray},
  stringstyle=\color{stringgreen},
  basicstyle=\ttfamily\footnotesize,
  breakatwhitespace=false,
  breaklines=true,
  numbers=none,
  captionpos=b,
  frame=lines,
  keepspaces=true,
  numbers=left,
  numbersep=5pt,
  showspaces=false,
  showstringspaces=false,
  showtabs=false,
  tabsize=1,
  columns=fullflexible 
}
\begin{document}

\title{A parameter study for LLL and BKZ with application to shortest vector problems}


\author{\IEEEauthorblockN{Tobias Köppl\IEEEauthorrefmark{1}, René Zander\IEEEauthorrefmark{2}, Louis Henkel\IEEEauthorrefmark{3}, Nikolay Tcholtchev\IEEEauthorrefmark{4}}
\IEEEauthorblockA{\IEEEauthorrefmark{1}Fraunhofer Institute FOKUS\\
Berlin, Germany\\
Email: tobias.koeppl@fokus.fraunhofer.de}
\IEEEauthorblockA{\IEEEauthorrefmark{2}Fraunhofer Institute FOKUS\\
Berlin, Germany\\
Email: rene.zander@fokus.fraunhofer.de}
\IEEEauthorblockA{\IEEEauthorrefmark{3}Fraunhofer Institute FOKUS\\
Berlin, Germany\\
Email: louis.henkel@fokus.fraunhofer.de}
\IEEEauthorblockA{\IEEEauthorrefmark{4}RheinMain University of Applied Sciences \& Fraunhofer Institute FOKUS\\
Berlin, Germany\\
Email: nikolay.tcholtchev@fokus.fraunhofer.de}
}

\maketitle
\begin{abstract}
In this work, we study the solution of shortest vector problems (SVPs) arising in terms of learning with error problems (LWEs). LWEs are linear systems of equations over a modular ring, where a perturbation vector is added to the right-hand side. This type of problem is of great interest, since LWEs have to be solved in order to be able to break lattice-based cryptosystems as the Module-Lattice-Based Key-Encapsulation Mechanism published by NIST in 2024. Due to this fact, several classical and quantum-based algorithms have been studied to solve SVPs. Two well-known algorithms that can be used to simplify a given SVP are the Lenstra–Lenstra–Lovász (LLL) algorithm and the Block Korkine–Zolotarev (BKZ) algorithm. LLL and BKZ construct bases that can be used to compute or approximate solutions of the SVP. We study the performance of both algorithms for SVPs with different sizes and modular rings. Thereby, application of LLL or BKZ to a given SVP is considered to be successful if they produce bases containing a solution vector of the SVP.
\end{abstract}
\ \\ \\

\section{Introduction}

Due to significant progress that has been made to improve both quantum hardware and algorithms, new concepts for public-key cryptosystems have to be developed. The goal is to design cryptosystems or cryptographic algorithms that are resistant to attacks using quantum computers. In literature, such algorithms are denoted as Post-quantum cryptography (PQC) algorithms \cite{kumar2020post,bernstein17}. 

One of the different PQC algorithms that have been investigated so far is the key encapsulation algorithm described in the NIST report FIPS 203 \cite{NIST203}. This algorithm is based on a mathematical problem given by a linear system of equations over a modular ring, where he matrix of this system is sampled from a uniform distribution. The right-hand side and matrix of this linear system of equations form the public key of the key encapsulation algorithm. However, in addition to that, a perturbation or error vector is sampled from a Gaussian distribution and added to the right-hand side of the linear system of equations. Depending on the standard deviation, the absolute values within the error vectors are usually bounded by $2$ or $3$. As a consequence, the public key cannot be used to recover the secret key, which is the solution of the overall system. This type of problem is known as learning with errors problem (LWE problem) \cite{Albrecht17}. 

Due to the significance of LWE problems several solution strategies have been developed \cite{Peikert16}. One of them is to reformulate the LWE problem as an optimization problem whose solution vector consists of the secret key and the error vector. Moreover, it is assumed that both vectors have components with small absolute values. For this reason, those problems are referred to as short vector problems (SVPs) \cite{uemura2021shortest}. Contrary to an LWE problem, the solutions of a SVP are given by vectors over the ring of integers. This facilitates the design of solution algorithms, since in terms of a SVP no modular arithmetics is required anymore. It is sufficient to use standard arithmetics in integer rings. 

However, up to now, there are no efficient algorithms for solving SVPs of relevant size. To facilitate the solution of an SVP, algorithms like the Lenstra–Lenstra–Lovász basis reduction algorithm (LLL algorithm) or the Block Korkine–Zolotarev algorithm (BKZ algorithm) are utilized \cite{Schneider10,Xia22,LLL82,koppl2024resilience,li2025complete}. These algorithms consider the solutions of SVPs as elements of a lattice and transform the basis of this lattice such that the new basis is almost orthogonal and consisting of short vectors. By means of such a basis, the solution of a SVP can be computed or approximated in a more convenient way. In some cases, the basis can contain the solution of the SVP itself. Quite often LLL and BKZ are combined with other solvers for discrete optimization problems \cite{prokop2024grover,Lv22}.

In this work, we apply the LLL and BKZ to LWE problems of different sizes. In addition to that the modulus of the underlying modular rings is varied. For each setting, the probability of solving the corresponding SVP is computed. Our goal is to determine for which problem sizes and moduli lattice reduction algorithms can solve a given SVP. As a consequence one can derive criteria, which can be utilized to decide when it is required to combine lattice reduction algorithms with further solvers. The remainder of this work is organized as follows: In Section \ref{sec:SVPs}, we describe the fundamentals of lattice-based cryptography and SVPs. The following section contains a short description of the LLL and BKZ algorithm. In Section \ref{sec:Experiments}, the results of our numerical tests are summarized. The paper is concluded by a short outlook.

\section{Lattice-based cryptography and SVPs}
\label{sec:SVPs}

In this section, we recall the fundamentals of the key generator described in the NIST report FIPS 203 \cite{NIST203}. Since this cryptosystem is a symmetric public key system, the key generator has to produce both a private and a public key for encryption and decryption. The public key consists of vectors 
$$
\hat{t}_l \in \mathbb{Z}_q^n, l \in \left\{1,\ldots,k\right\},
$$
which can be compiled to a single vector $t \in \mathbb{Z}_q^{nk}$:
$$
t = \left( \hat{t}_1,\ldots, \hat{t}_k \right)^T
$$
and further vectors
$$
a^{(ij)} = \left(a_1^{(ij)},\ldots,a_n^{(ij)} \right)^T \in \mathbb{Z}_q^{n}, i,j \in \left\{1,\ldots,k\right\},
$$
where $n \in \mathbb{N}$ is the characteristic key length of the public key. For $k$ it holds according to \cite{NIST203}: $k \in \left\{2,3,4\right\}$ and $q \in \mathbb{N}$ is the modulus of the ring $\mathbb{Z}_q$. The vectors $a^{(ij)}$ are sampled from a uniform distribution. Based on $a^{(ij)}$, we construct matrices $A_{ij} \in \mathbb{Z}_q^{n \times n}$:
$$
A_{ij} = 
\begin{pmatrix} a_1^{(ij)} & -a_{n}^{(ij)} & \ldots & -a_2^{(ij)} \\ 
                a_2^{(ij)} &  a_1^{(ij)}     & \ldots & -a_3^{(ij)} \\
                \vdots     &  \vdots         &   & \vdots \\
                a_{n}^{(ij)} & a_{n-1}^{(ij)}  & \ldots & a_{1}^{(ij)}
\end{pmatrix}
$$
The matrix $A_{ij}$ is a T\"oplitz matrix, i.e., on each diagonal from top to bottom is the same element \cite{Gray06}. Each column vector is generated by a cyclic permutation of the components in $a^{(ij)}$ and to the entries in the upper triangular part a minus sign is assigned. All the submatrices can assigned to a further matrix $A \in \mathbb{Z}_q^{nk \times nk}$:
$$
A= \begin{pmatrix} A_{11} & \cdots & A_{1k} \\ \vdots & \ddots &  \vdots \\  A_{k1} & \cdots & A_{kk}\end{pmatrix}.
$$
Next, we sample the secret vector or private key $s \in \mathbb{Z}_q^{nk}$ as well as the error vector $e \in \mathbb{Z}_q^{nk}$ from a discrete Gaussian distribution such that the absolute values of $s$ and $e$ are bounded by $2$ or $3$. $s$ and $e$ have the following structure:
\begin{align*}
s &= \left( \hat{s}_1, \ldots, \hat{s}_k \right)^T,\;\hat{s}_l \in \mathbb{Z}_q^{n}, \\
e &= \left( \hat{e}_1, \ldots, \hat{e}_k \right)^T,\;\hat{e}_l \in \mathbb{Z}_q^{n}.
\end{align*}
Using $A$, $s$ and $e$, the missing part of the public key $t$ can be determined:
\begin{equation}
    \label{eq:LWE1}
t = A \cdot s + e,
\end{equation}
where all the arithmetic operations have to be performed in the ring $\mathbb{Z_q}$. For a potential attacker, the matrix $A$ and the vector $t$ are known, while the vectors $s$ and $e$ have to be determined. This problem is know as LWE problem. Since a LWE problem is not well-posed, \eqref{eq:LWE1} is reformulated. In the remainder of this section, we present the most important steps. $\mathbf{0}$ denotes either a vector filled with zeros or a matrix filled with zeros.
\ \\
\begin{itemize}
    \item[]\textbf{Step (1):} In a first step the lattice 
    $$
    \Lambda = \left\{ \left. x \in \mathbb{Z}^{2m+1} \right| x^T \left( \left. A \right| \left. I_m \right| -t \right) \equiv \mathbf{0} \mod q \right\},
    $$
    is constructed, where $I_m$ is the identity matrix with $m$ rows and columns. Obviously, it holds that the vector
    $$
    y = \left( \left. s \right| \left. e \right| 1 \right)^T \in \mathbb{Z}^{2m+1}
    $$
    is contained in $\Lambda$. For convenience, we set: $m=kn$
    \ \\ 
    \item[]\textbf{Step (2):} The rows of the matrix 
    \begin{equation}
    \label{eq:LWE2}
    B = \begin{pmatrix} I_m & -A^T & \mathbf{0} \\ \mathbf{0} & q I_m & \mathbf{0} \\  \mathbf{0} & t^T & 1
    \end{pmatrix}
    \end{equation}
    form a basis of the lattice $\Lambda$, provided that computing linear combinations of the rows integer valued coefficients are used. Compiling the matrix $B$ in this way is also known as Bai and Galbraith’s embedding \cite{Albrecht17}. 
    \ \\
    \item[]\textbf{Step (3):} We compute $y$ by solving the following optimization problem (shortest vector problem, SVP):
    \begin{equation}
    \label{eq:LWE3}
    z = \text{argmin}_{x \in \mathbb{Z}^{2m+1} \setminus \left\{\mathbf{0}\right\}} \left\| x^T B \right\|_2.
    \end{equation}
    After that, we set:
    $$
    z^T B = y^T \Leftrightarrow y = B^T \cdot z.
    $$
    Solving an SVP is based on the fact that $s$ and $e$ are sampled from a discrete Gaussian distribution with a small standard deviation. 
\end{itemize}

\section{Lattice reduction algorithms}

In this section, we briefly discuss classical methods for lattice reduction. Let $$\Lambda(B)=\left\{\sum_{i=1}^nx_ib_i\mid x_i\in\Z\right\}$$ be a lattice with basis $B=(b_1,\dotsc,b_n)$ for linearly independent vectors $b_i\in\R^d$ for $n\leq d$. Here, we consider the full-dimensional case $n=d$. Such bases are not unique as lattices are invariant under unimodular transformations, i.e., $\Lambda(B)=\Lambda(BU)$ for all $U\in\mathrm{GL}_n(\Z)$ with $\det(U)=\pm1$. In particular, every lattice of dimension $n>1$ has infinitely many bases. We denote the length of a shortest vector of the lattice $\Lambda$ by $\lambda_1(\Lambda)$.

The $\gamma$-SVP problem asks for finding an approximate shortest non-zero vector in the lattice, i.e., a vector $v\in\Lambda\setminus\{0\}$ with $\|v\|\leq\gamma\lambda_1(\Lambda)$, where $\gamma\geq1$ is the approximation factor. Breaking an LWE based cryptosystem requires the ability of finding a shortest vector up to an approximation factor that is linear in the lattice dimension \cite{Peikert16}. 

Classical lattice reduction algorithms such as Lenstra-Lenstra-Lovász (LLL) \cite{LLL82} and Block Korkin-Zolotarev (BKZ) \cite{Schnorr94} and variants thereof can be applied to obtain a basis $B'=(b_1',\dotsc,b_n')$ of relatively short and nearly orthogonal vectors. Here, ``relatively short'' means that the resulting bases are guaranteed to contain a vector of length $\gamma\cdot\lambda_1(\Lambda)$ up to an approximation factor $\gamma\in\R$ \cite{Schneider10}.

The LLL algorithm \cite{LLL82} has a polynomial runtime with respect to the lattice dimension $n$, and is guaranteed to solve $\gamma$-SVP for approximation factors exponential in the lattice dimension $n$: It is parametrized by $\delta\in(1/4,1]$ which appears in the Lovász condition that controls the strictness of the reduction process, i.e. how short and orthogonal the vectors of the reduced basis are. Thereby, it balances the trade-off between computational efficiency and the quality of the reduced basis.
The LLL algorithm produces a basis whose first vector satisfies
$$ \|b_1\|\leq (\delta-1/4)^{\frac{1-n}{2}}\lambda_1(\Lambda) $$
In practice, the LLL algorithm finds much shorter vectors than guaranteed by this worse-case bound \cite{Schnorr94}.

The BKZ algorithm \cite{Schnorr94} combines enumeration based methods with LLL, and is parametrized by the block size parameter $\beta$ and the LLL parameter $\delta$. It produces bases whose first vector satisfies 
$$ \|b_1\|\leq (\gamma_{\beta})^{\frac{n-1}{\beta-1}}\lambda_1(\Lambda) $$
where $\gamma_{\beta}$ is the Hermite constant in dimension $\beta$ \cite{Schnorr94}.
The running time scales exponentially in the bock size $\beta$ for the enumeration step.
Therefore, it offers a trade-off between the running time and the quality of the output basis.

\section{Experiments}
\label{sec:Experiments}

This section shows experiments for solving LWE instances with LLL and BKZ lattice reduction algorithms. The LWE instances were generated based on the NIST FIPS 203 standard with varying parameters. Depending on the security parameter, the NIST standard uses block matrices of size $k=2,3,4$, with each block sized $n = 256$, and modulus $q=3329$. For simplicity, we will only consider block matrices of size $k=1$ but will vary the block size $n$ and modulus $q$. The secret vector $s$ and error vector $e$ in $\{-2,\dotsc,2\}^n$ are drawn from a centered binomial distribution with mean $0$. The LWE instances are transformed into shortest vector problems, and the aforementioned lattice reductions techniques are applied.

The experiments were performed on a dual Intel Xeon E5-2695 (2.10 GHz) 18 core processor.

\subsection{Results for LLL algorithm}
\label{sec:Experiments:LLL}

\begin{figure}[htbp]
\includegraphics[scale=1.1]{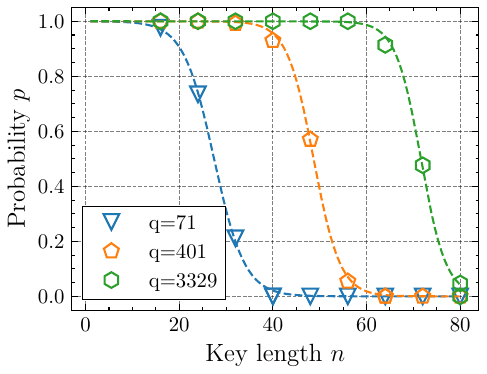}
\caption{The probability that the LLL algorithm recovers the secret for varying key length $n$ and modulus $q\in\{71,401,3329\}$. The curves were fitted using a sigmoid function $p_{\rho,\sigma}(n)=1-(1+\exp(\rho-\sigma n))^{-1}$ with parameters $\rho=9.32$, $\sigma=0.35$ for $q=71$, $\rho=17.00$, $\sigma=0.36$ for $q=401$, and $\rho=26.94$, $\sigma=0.38$ for $q=3329$.}
\label{fig:lll_1}
\end{figure}

For the experiments with the LLL algorithm we utilize the implementation in \texttt{fpylll} \cite{FPYLLL}. We choose the LLL parameter $\delta=0.99$ for all experiments. Experiments were conducted for security parameters $n=8k$ for $k=1,\dotsc,10$, and $q\in\{3,17,71,277,401,521,1031,3329\}$. For each combination of parameters $(n, q)$, $128$ LWE instances were generated. Based on these 128 instances, the probability that the LLL algorithm recovers the secret was calculated. For $q\in\{71,401,3329\}$ the results are shown in Figure \ref{fig:lll_1}. The results indicate that 1) for fixed modulus $q$ the success probability decreases exponentially with the key length $n$, and 2) for fixed key length $n$ the success probability grows with increasing modulus $q$. The first observation is as expected, as the search space grows exponentially with larger key size $n$.

\begin{figure}[htbp]
\includegraphics[scale=1.05]{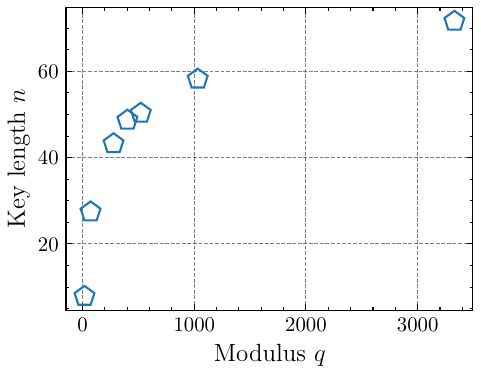}
\caption{The maximum key length $n$ for which the probability that the LLL algorithm recovers the secret is at least $50\%$ for varying modulus $q$. The values $n$ are computed as the ratio of the fit parameters $\rho/\sigma$.}
\label{fig:lll_2}
\end{figure}

The latter phenomenon is further studied in Figure \ref{fig:lll_2}. This Figure shows the maximum key length $n$ for which the probability that the LLL algorithm recovers the secret is at least $50\%$ for varying modulus $q$. The results indicate that this key length $n$ grows logarithmically with the modulus $q$. While this observation may be counter intuitive, recall that the size of the search space for the underlying LWE instance does only depend on key size $n$ and not on the modulus $q$ since $s,e\in\{-2,\dotsc,2\}^n$. The observations are further consistent with results in \cite{Laine15} establishing that success can be guaranteed with high probability for narrow error distributions and $\log_2 q>2n$. Similar to \cite{Laine15}, the results show that, in practice, significantly smaller moduli $q$ are vulnerable.

\subsection{Results for BKZ algorithm}
\label{sec:Experiments:BKZ}

\begin{figure}[htbp]
\includegraphics[scale=0.615]{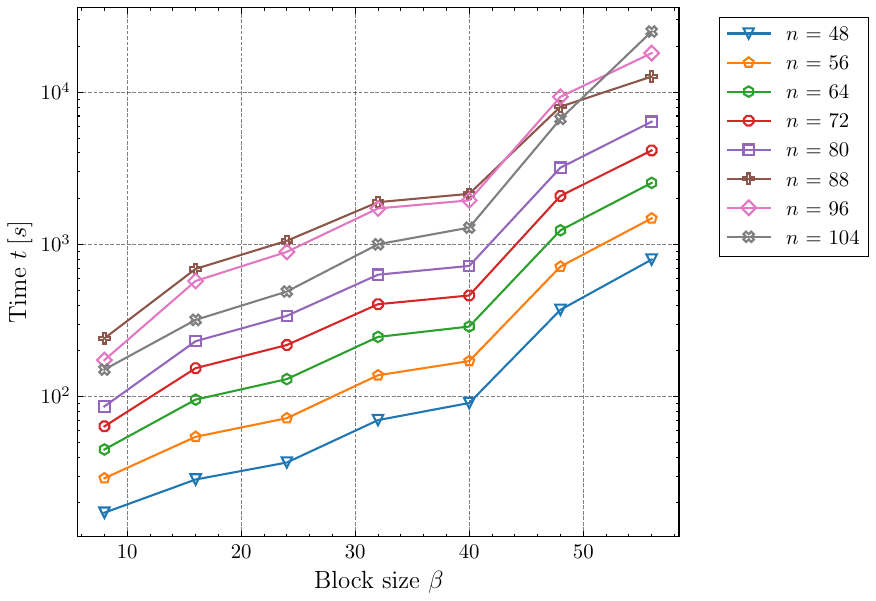}
\caption{The average runtime for the BKZ algorithm for varying block size $\beta$ and key size $n$, and modulus $q=401$.}
\label{fig:bkz_runtime}
\end{figure}

\begin{figure*}[htbp]
\centering
\subfigure[Modulus $q=71$]{
    \label{fig:bkz_71}
    \includegraphics[scale=0.95]{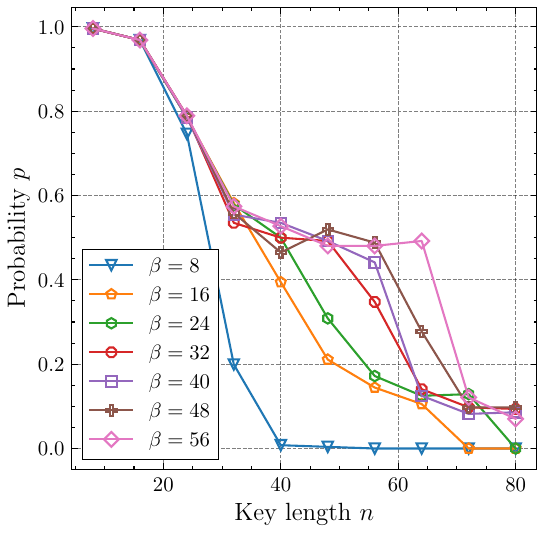}
}
\subfigure[Modulus $q=401$]{
    \label{fig:bkz_401}
    \includegraphics[scale=0.95]{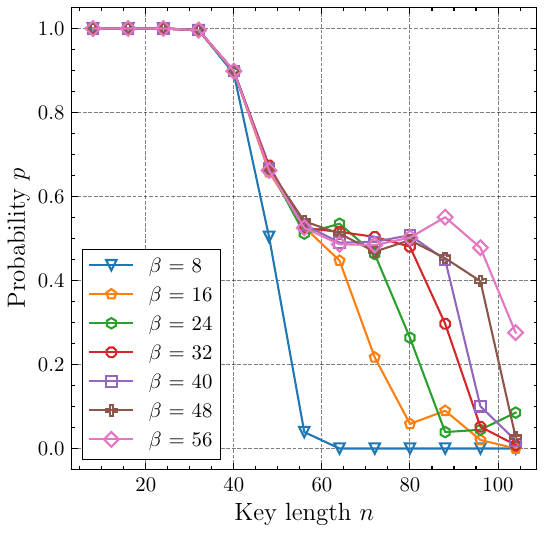}
}

\subfigure[Modulus $q=3329$]{
    \label{fig:bkz_3329}
    \includegraphics[scale=0.95]{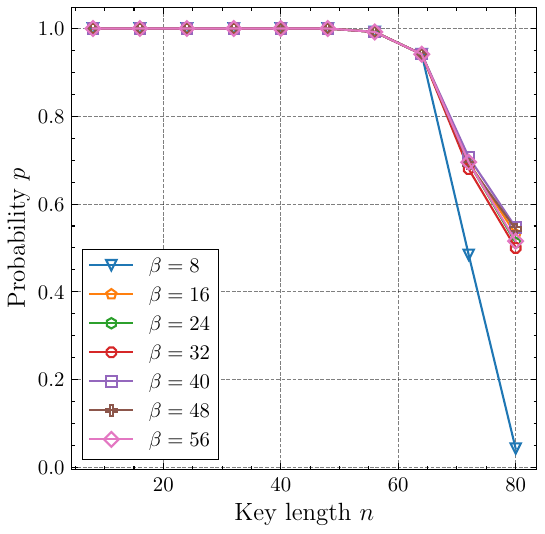}
}

\caption{The probability that the BKZ algorithm recovers the secret for varying key length $n$ and block size $\beta$, and moduli $q=71$ ((a) top left), $q=401$ ((b) top right), and $q=3329$ ((c) bottom).}
\label{fig:bkz}

\end{figure*}

For the experiments with the BKZ algorithm we utilize the implementation of \texttt{BKZ 2.0} in \texttt{fpylll}. We choose the LLL parameter $\delta=0.99$ for all experiments. Experiments were conducted for security parameters $n=8k$ for $k=1,\dotsc,14$, and $q\in\{71,401,3329\}$, and for varying block size $\beta=8l$ for $l=1,\dotsc,7$. For each key length $n\leq 88$, $512$ LWE instances were generated, and for each $n>88$, the sample size was reduced to $256$ LWE instances due to the larger computational costs.  
Additionally, for parameters $(n,\beta)$ with $n>88$, the sample size was further reduced. Instances that encountered errors due to insufficient floating-point precision during the BKZ calculations were discarded. For $n=112$ and $\beta\geq16$ the BKZ algorithm failed in almost all cases. For each $(n,\beta)$ the success probability of the BKZ algorithm was calculated. The results are shown in Figure \ref{fig:bkz}. As expected, we observe that the success probability for fixed key length $n$ increases with the block size $\beta$. 
Notably, for $\beta=56$ the success probability for relatively large keys $n=104$ ($q=401$) is still at $28\%$. 
Similar to the experiments utilizing the LLL algorithm, we notice that for fixed key length $n$, the success probability is higher for larger modulus $q$.
Moreover, the average runtime of the BKZ algorithm is shown in Figure \ref{fig:bkz_runtime} in a logarithmic plot. As expected, it shows that the average runtime grows exponentially with the bock size $\beta$.


\subsection{Conclusions and implications for lattice-based cryptography}

\begin{itemize}
    \item The results presented is this study underscore the importance of an appropriate choice of security parameters in latticed-based cryptography. In particular, the findings in Section \ref{sec:Experiments:LLL} indicate that an LWE instance can be solved by LLL even for large key size $n$ if the modulus $q$ is chosen too large. A similar behavior was observed for the BKZ algorithm.

    \item There are only a few open-source libraries available for lattice reduction algorithms. Although \texttt{fpylll} receives regular updates, its documentation remains insufficiently detailed and the selection of hyperparameters is particularly unclear, leading to potential errors and making it difficult to compare results. 

    \item Related work providing theoretical and experimental analysis of the cost for solving SVPs with lattice reduction algorithms is given by \cite{Albrecht17},\cite{Laine15},\cite{Schneider10}.  Essentially, these works share the conclusions that: 1) Empirically observed results are better than what is expected form theoretical analysis. 2) There is an insufficient availability of state-of-the art software libraries for testing attacks on SVPs. 
\end{itemize}

\section{Outlook}

While there is a growing number of open-source libraries providing implementations of PQC algorithms, very few libraries providing the tools for empirically testing the resilience of PQC implementations exist to date. The latter would however be necessary for further gaining confidence in the security of these new methods for which there is less experience regarding implementations and cryptanalysis compared to established schemes. Accordingly, in a joint statement from partners from 18 EU member states it is recommended to ``deploy PQC in hybrid
solutions for most use-cases, i.e. combining a deployed cryptographic scheme with PQC in such a
way that the combination remains secure even if one of its components is broken'' \cite{BSI}.

Lattice-based cryptography can be challenged by a variety of attackers, e.g., classical lattice reduction algorithms (LLL, BKZ) and quantum-accelerated versions thereof \cite{prokop2024grover}, hybrid quantum-classical optimization algorithms (QAOA) \cite{koppl2024resilience}, or potentially emerging new quantum algorithms. Robust testing of lattice-based cryptography requires suitable software libraries that are optimized for integrated classical/quantum high-performance computing. 
The open-source high-level quantum programming framework Eclipse Qrisp \cite{Qrisp} -- with its recent integration of the high-performance computing library JAX -- bridges the gap for enabling researchers to conduct experiments using hybrid classical/quantum high-performance computing in the era of fault-tolerant quantum computing.

\section*{Acknowledgment}
This work was supported by the EU Horizon Europe Framework Program under Grant Agreement no 101119547 (PQ-REACT).
In terms of PQ-REACT methods and tools for benchmarking PQC algorithms are investigated. In particular, testing the resilience of PQC algorithms against various attacks is a central pillar and part of ongoing work in the PQ-REACT project.

\end{document}